\documentclass[letterpaper, 10 pt, conference]{ieeeconf}  

\IEEEoverridecommandlockouts                              

\usepackage{color}

\usepackage{times,amsmath,epsfig}
\usepackage{amssymb}
\usepackage{cite}

\usepackage{algorithmic}
\usepackage{algorithm}
\usepackage{url}

\usepackage[font=small,format=plain,up,up]{caption}
\usepackage{subcaption}
\usepackage{needspace}

\usepackage{tikz}
\usetikzlibrary{shapes,arrows}
\input{mysymbol.sty}
\tikzstyle{phantom vertex} = [ ellipse, 
                               anchor = center, 
                               minimum height = 1*\unit, 
                               minimum width  = 1*\unit,
                               inner sep=0pt,
                               anchor=center]
\tikzstyle{red vertex}   = [black, fill = red!20,   phantom vertex, draw]
\tikzstyle{black vertex} = [black, fill = black!20, phantom vertex, draw]
\tikzstyle{blue vertex}  = [black, fill = blue!20,  phantom vertex, draw]
\tikzstyle{green vertex} = [black, fill = green!20,  phantom vertex, draw]
\tikzstyle{yellow vertex} = [black, fill = yellow!20,  phantom vertex, draw]
\tikzstyle{cyan vertex} = [black, fill = cyan!20,  phantom vertex, draw]
\tikzstyle{vertex}       = [draw, phantom vertex]

\tikzstyle{point} = [ellipse, inner sep=0pt, draw, fill=white, anchor = center,
                     minimum height = 0.05*\unit, minimum width  = 0.05*\unit]

\renewcommand{\baselinestretch}{0.98}

\newtheorem{mylemma}{\hspace{-11pt}\bf Lemma}
\newtheorem{myproposition}{\hspace{-11pt}\bf Proposition}

\title{\LARGE \bf
Network Inference from Consensus Dynamics
}

\author{Santiago Segarra, Michael T. Schaub, Ali Jadbabaie
\thanks{{The authors are with the Institute for Data, Systems, and Society, Massachusetts Institute of Technology. A. Jadbabaie's research is supported by a Vannevar Bush fellowship from the Office of Secretary of Defense. Emails: \{segarra, mschaub, jadbabai\}@mit.edu.}}}

\begin{document}

\maketitle
\thispagestyle{empty}
\pagestyle{empty}

\begin{abstract}
We consider the problem of identifying the topology of a weighted, undirected network $\mathcal G$ from observing snapshots of multiple independent consensus dynamics. Specifically, we observe the opinion profiles of a group of agents for a set of $M$ independent topics and our goal is to recover the precise relationships between the agents, as specified by the unknown network $\mathcal G$. In order to overcome the under-determinacy of the problem at hand, we leverage concepts from spectral graph theory and convex optimization to unveil the underlying network structure. More precisely, we formulate the network inference problem as a convex optimization that seeks to endow the network with certain desired properties -- such as sparsity -- while being consistent with the spectral information extracted from the observed opinions. This is complemented with theoretical results proving consistency as the number $M$ of topics grows large. We further illustrate our method by numerical experiments, which showcase the effectiveness of the technique in recovering synthetic and real-world networks.
\end{abstract}


\section{INTRODUCTION}\label{S:intro}

The study of networks and multi-agent systems has attracted enormous interest over the last years.
Network-based problem formulations abound in diverse application domains, ranging from socio-economical to biological settings, and from physical to engineering systems~\cite{Strogatz2001,Boccaletti2006,Newman2010}.
Often, these systems display a rich dynamical behavior that emerges from an interplay of the non-trivial connection topology of the network.
In this context, consensus has been one of the most popular and well-studied dynamics on networks~\cite{Jadbabaie2003,Olfati-Saber2007,Ren2005}.
This is due to both its analytic tractability as well as its simplicity in approximating several fundamental behaviors.
For instance, in socio-economic domains consensus provides a model for opinion formation in a society of individuals.  
For engineering systems, it has been considered as a basic building block  for an efficient  distributed computation  of  global functions in networks of sensors, robots, or other~agents.  

However, especially in the biological and social domains, the true couplings between the agents are usually unknown, and have to be inferred from data~\cite{D’Haeseleer2000,Timme2014,Brugere2016}.
Network inference, though often not discussed explicitly, is thus a fundamental constituent of network analysis.
Different notions of network inference have been considered in the literature, with examples ranging from the estimation of `functional' couplings based on pairwise statistical association measures (correlation, mutual information), to causal inference~\cite{Spirtes2016}.
In this paper, we are interested in what is often called structural or topological inference~\cite{Timme2014,Brugere2016}: given a system of dynamical units, we want to infer their direct `physical' interactions, e.g., for a distributed system on a network, we want to infer the exact weighted adjacency relationships between the nodes.

Optimization-based strategies for such inference tasks have been proposed in the literature\cite{Julius2009,Candes2008}.
In relation to the inference of networks from consensus-like dynamics, \cite{Shahrampour2013} and \cite{Shahrampour2015} consider inferring the network based on observing the (cross-)power spectral density of a consensus process driven by noise, and a node knockout strategy.
Related approaches have also been pursued in~\cite{Materassi2012,Nabi-Abdolyousefi2012}.
Further methods combining notions from spectral identification with optimization techniques are considered in~\cite{Hayden2016a,Hayden2016,Yuan2011}.
Moreover, \cite{Mauroy2017} proposes an interesting approach for the identification of nonlinear systems based on the observation of a few nodes only, however the recovery is limited to spectral properties of the underlying network structure, rather than the full topology. 
In our previous work~\cite{Segarra2016, segarra_topology_ssp}, we studied how independent steady-state observations of the same linear network process can be used to extract information about the eigenvectors of the unknown underlying network. 

Departing from the existing literature, in this paper we make very few assumptions on the unknown structure and the samples obtained from observing the consensus process. 
In particular, we neither impose a specific sampling scheme in which samples are drawn at known time instances, nor do we assume that we have access to a complete time-series of observations.
Instead, we are given a statistical characterization of the initial inputs, and observe a set of $M$ independent snapshots of the responses to such inputs after some unknown time intervals.
Surprisingly, even under these mild assumptions it is possible to (approximately) infer the network structure, as we demonstrate in the sequel.

After a brief review of preliminary concepts in Section~\ref{S:preliminaries}, the recovery problem is described rigorously in Section~\ref{S:problem_formulation}. 
Section~\ref{S:network_topology_inference} discusses how several spectral properties of the unknown network can be inferred from the observation of a set of output signals (Sections~\ref{Ss:inferring_eigenvectors} and \ref{Ss:inferring_eigenvalues}) and then proposes a convex optimization formulation for the recovery problem that takes into account this spectral information (Section~\ref{Ss:optimal_laplacian}).
We illustrate the effectiveness of this approach through a series of numerical experiments on synthetic as well as real-world networks in Section~\ref{S:numerical_experiments}.
A brief outlook in Section~\ref{S:conclusion} identifying extensions and future lines of work wraps-up the paper.

\noindent
{\bf Notation:} The entries of matrix $\mathbf{X}$ and (column) vector $\mathbf{x}$ are denoted by $X_{ij}$ and $x_i$, respectively; to avoid confusion, the alternative notation $[\bbX]_{ij}$ and $[\bbx]_i$ will be used occasionally, when dealing with indexed families of matrices (vectors). The notation $^\top$, $\mathbb{E}(\cdot)$, and $\mathbb{P}(\cdot)$ stands for transpose, expected value, and probability, respectively; $\mathbf{0}$, $\mathbf{1}$, and $\bbI$ refer to the all-zero vector, the all-one vector, and the identity matrix, where the sizes are clear from context. For a vector $\bbx$, $\diag(\mathbf{x})$ is a diagonal matrix whose $i$th diagonal entry is~$x_i$. 

\section{PRELIMINARIES}\label{S:preliminaries}

{\bf Networks.} A weighted and undirected network $\ccalG$ consists of a node set $\ccalN$ of known cardinality $N$, an edge set $\ccalE$ of unordered pairs of elements in $\ccalN$, and non-negative edge weights $A_{ij}\in\reals_+$ such that $A_{ij}=A_{ji}\neq 0$ for all $(i,j)\in\ccalE$. The neighborhood of $i$ is defined as the set of nodes $\mathcal{N}_i = \{j \,| \, (j,i)\in\mathcal{E}\}$ connected to $i$. The edge weights $A_{ij}$ can be conveniently collected as entries of the symmetric adjacency matrix $\bbA$. Defining the diagonal degree matrix $\bbD:=\diag(\bbA\bbone)$, the (combinatorial) Laplacian matrix associated with network $\ccalG$ is given by $\bbL:=\bbD-\bbA$ and can be shown to be positive semi-definite~\cite{Merris1994}. Since $\bbL$ is a symmetric and real matrix, it is diagonalized by a unitary matrix $\bbV$, i.e. $\bbL = \bbV \bbLambda \bbV^\top$, where the diagonal matrix $\bbLambda := \diag(\bblambda)$ contains the eigenvalues of~$\bbL$. Throughout the paper,
we are going to assume that all eigenvalues of $\bbL$, $0 = \lambda_1 < \lambda_2 < \ldots < \lambda_N$, are unique (non-degenerate). This assumption is not fundamental from a technical viewpoint, but simplifies the presentation of our results.
In particular, this implies that $\ccalG$ is connected. 
The Laplacian matrix $\bbL$ can be used to model local dynamics on the associated network $\ccalG$, as discussed next.

\vspace{0.05cm}
{\bf Discrete-time consensus dynamics.} Define the state vector $\bbx \in \reals^N$ where the value $x_i$ corresponds to the opinion of agent $i$ in the network.
We consider a discrete-time linear consensus dynamics~\cite{Ren2005,Olfati-Saber2007,Jadbabaie2003} in which $\bbx$ evolves locally, i.e., the value of $x_i$ at a given instant depends exclusively on the previous values of $\bbx$ at node $i$ and its neighborhood $\mathcal{N}_i$:
\begin{equation}\label{E:consensus_node_1}
x_i[t] = x_i[t-1] + \alpha_{t} \sum_{j \in \ccalN_i} A_{ij} (x_j[t-1] - x_i[t-1] ),
\end{equation}
where $t = 1, 2, \ldots$ indicates discrete time instants. 
According to \eqref{E:consensus_node_1}, agent $i$ updates its state as a linear combination of its own state in the previous time step and the weighted discrepancy with its neighbors in the previous time step.
In this context, $\alpha_{t}$ indicates the relative weight that node $i$ gives to this discrepancy in the respective update (the `learning rate' of the nodes at time $t$).
From the definition of $\bbL$ it readily follows that \eqref{E:consensus_node_1} can be expressed in vector form as $\bbx [t] = \bbx[t-1] - \alpha_{t} \bbL \bbx[t-1]$, or more compactly:
\begin{equation}\label{E:consensus_all_nodes}
\bbx[t] = (\bbI - \alpha_{t} \bbL) \bbx[t-1].
\end{equation}
We refer to the above dynamics as discrete-time consensus because, under mild conditions on $\alpha_{t}$, the state of all agents coincides asymptotically, i.e. $\bbx[t] = \beta {\bf 1}$ for $t \rightarrow \infty$, where $\beta \in \mathbb R$ is a constant.

Throughout the paper, we will use $\bbx$ to refer to the initial signal, i.e. $\bbx := \bbx[0]$, and we denote by $\bby$ the observation of the dynamics at a specific time $T$ of interest, i.e., $\bby := \bbx[T]$. It follows then that $\bby$ is related to $\bbx$ via
\begin{equation}\label{E:final_single_consensus}
\bby = \prod_{t=1}^T (\bbI - \alpha_t \bbL) \, \bbx.
\end{equation}

\vspace{0.05cm}
{\bf Sub-exponential random variables.} A random variable $x$ with mean $\mu = \mathbb{E}(x)$ is sub-exponential if there exist non-negative parameters $(\nu, b)$ such that, for all $\gamma$ satisfying $|\gamma| <1/b$, it holds that~\cite{Rigollet2015,Boucheron2013}
\begin{equation}\label{E:def_sub_exponential}
\mathbb{E}(e^{\gamma(x-\mu)}) \leq \exp \left ({\frac{\nu^2\gamma^2}{2}} \right).
\end{equation}
When $x$ is sub-exponential, from the classical Chernoff bound one can derive that~\cite{Rigollet2015,Boucheron2013}
\begin{equation}\label{E:def_sub_exponential_bound}
\mathbb{P}(|x - \mu | \geq l) \leq 
\begin{cases}
2\exp \big({-\frac{l^2}{2\nu^2}}\big) & \text{if $0 \leq l \leq \nu^2/b$},\\
2\exp \left({-\frac{l}{2b}}\right) & \text{if $l>\nu^2/b$}.
\end{cases}
\end{equation}
Moreover, given two sub-exponential random variables $x$ and $y$ with corresponding parameters $(\nu_1, b_1)$ and $(\nu_2, b_2)$, the sum $z = x+y$ is also sub-exponential with parameters $(\sqrt{\nu_1^2 + \nu_2^2}, \max(b_1, b_2))$.

\section{PROBLEM FORMULATION}\label{S:problem_formulation}
To motivate our problem setup, let us consider the context of social networks.
Assume that we observe, at a specific instant in time, the opinion profile of all agents in a network $\mathcal G$ regarding $M$ independent topics, each of which evolved according to a consensus dynamics like the one in \eqref{E:final_single_consensus}.
The discussion about each of the $M$ topics, which we index via $k \in \{1,\ldots, M\}$, may have started at a different point in time -- corresponding to unknown durations $T_k$ for each topic.
Furthermore, the interactions between the agents may have been heterogeneous across time and topics -- associated with unknown learning rates $\alpha_{t}^{(k)}$. 
Our goal is now to recover the underlying social network $\bbL$ from the observation of $M$ opinion profiles $\{\bby_k\}$ at a given time instant.

Formally, consider $M$ different consensus dynamics evolving on a single network encoded by the Laplacian $\bbL$.
Each dynamics corresponds to a distinct input $\bbx_k$ and diffusion rates $\{\alpha_t^{(k)}\}$ as in \eqref{E:final_single_consensus}. 
Our goal is to recover $\bbL$ from a single snapshot of the state vectors $\bby_k$ of the $M$ consensus dynamics. 
More precisely, we have access to one observation $\bby_k$ for each dynamics, where $\bby_k$ is given by
\begin{equation}\label{E:final_multiple_consensus}
    \bby_k = \prod_{t=1}^{T_k} (\bbI - \alpha_{t}^{(k)} \bbL) \, \bbx_k, \quad k = 1, \ldots, M.
\end{equation}
Notice that, while $\bbL$ is the unknown of interest, we do not assume that we know $T_k$, i.e. how long each consensus dynamics has been running, nor do we assume the knowledge of the diffusion rates $\alpha_{t}^{(k)}$. 
We assume merely that we have a statistical characterization of the unknown inputs $\bbx_k$, which we model here as independent realizations of a zero-mean multivariate normal random variable $\bbx \sim \ccalN( {\bf{0}}, \sigma^2 \bbI)$, of unknown power $\sigma^2$.

The above problem is markedly under-determined, thus requiring us to consider a novel recovery scheme for $\bbL$ that combines spectral information with a regularized convex optimization approach (see Section~\ref{S:network_topology_inference}).
To ensure that there \emph{is} some information about $\mathcal G$ contained in the snapshots $\{\bby_k\}$, we assume that the two following mild technical conditions hold. First, $T_k < \infty$ for all $k$, meaning that the dynamics has not reached consensus yet, in which case all information about the network structure would be completely lost.
Second, we assume that the deterministic rates $\alpha_t^{(k)}$ are small enough to ensure asymptotic convergence of \eqref{E:final_multiple_consensus}. 
Specifically, $0 < \alpha_{t}^{(k)} <1/\lambda_{\max}(\bbL)$ for all $k, t$.

\section{TOPOLOGY INFERENCE FROM CONSENSUS}\label{S:network_topology_inference}
The goal is to design a recovery scheme that is able to find $\bbL$, despite the scarce information about $\mathcal G$ at our disposal.
Our discussion is divided into three parts.
In Sections \ref{Ss:inferring_eigenvectors} and \ref{Ss:inferring_eigenvalues}, we highlight how information about the eigenvectors and eigenvalues of $\bbL$ can be inferred from the observation of $\{\bby_k\}$, and prove some asymptotic consistency results for the limit of large sample size $M$.
The guiding idea is the following: since the inputs $\bbx_k$ in \eqref{E:final_multiple_consensus} are realizations of \emph{white} Gaussian noise, the \emph{color} of the outputs $\bby_k$ must contain information about the unknown $\bbL$. 

In Section~\ref{Ss:optimal_laplacian}, we use these insights to design an inference method for $\bbL$, based on a regularized convex optimization problem.
More precisely, we propose to recover the unknown $\bbL$ by solving an optimization problem that searches for a valid Laplacian consistent with the spectral information extracted from the observations $\bby_k$, while promoting a desirable sparse structure.


\subsection{Inferring the eigenvectors}\label{Ss:inferring_eigenvectors}

Define the linear operators $\bbH_k := \prod_{t=1}^{T_k} (\bbI - \alpha_{t}^{(k)} \bbL)$ so that $\bby_k = \bbH_k \bbx_k$ for all $k$. Leveraging the eigendecomposition $\bbL = \bbV \bbLambda \bbV^\top$ we can write
\begin{align}\label{E:redefinition_filter_H}
    \bbH_k = \bbV \Big(\prod_{t=1}^{T_k} (\bbI - \alpha_{t}^{(k)} \bbLambda ) \Big) \bbV^\top = \bbV \bbLambda_k \bbV^\top,
\end{align}
where we have implicitly defined the diagonal matrices $\bbLambda_k := \prod_{t=1}^{T_k} (\bbI - \alpha_{t}^{(k)} \bbLambda )$ for all $k$. Notice that the diagonal entries of $\bbLambda_k$ are bounded between $(0,1]$. 
Consider the \emph{sample} covariance matrix
\begin{equation}\label{E:sample_covariance}
\bbS_M := \frac{1}{M} \sum_{k=1}^M \bby_k \bby_k^\top,
\end{equation}
and notice that for $k \neq k'$ the vectors $\bby_k$ and $\bby_{k'}$ are independent, but not identically distributed. More precisely, 
\begin{equation}\label{E:covariance_y_i}
\mathbb{E}(\bby_k \bby_k^\top) = \bbH_k \mathbb{E}(\bbx_k \bbx_k^\top) \bbH_k^\top = \sigma^2 \bbH_k^2,
\end{equation}
where we used that $\bbH_k$ is symmetric and deterministic, and that $\bbx_k$ is white.  
Note that, in general, $\bbS_M$ will not converge to the covariance of any specific $\bby_k$ for increasing $M$.
Hence the procedure based on spectral templates developed in~\cite{Segarra2016, segarra_topology_ssp} is not applicable here. 
However, as we demonstrate in the sequel, $\bbS_M$ can still be used to recover the eigenbasis $\bbV$ of the unknown Laplacian $\bbL$.

Before presenting our result formally in Proposition~\ref{P:diagonal}, we state the following lemma, which will be instrumental in the subsequent proof [cf. \eqref{E:def_sub_exponential}].

\begin{mylemma}\label{L:subexponential}
\emph{
Let $x \sim \ccalN(0, \sigma_1^2)$ and $y \sim \ccalN(0, \sigma_2^2)$ be independent zero-mean scalar Gaussian random variables. Then, the random variable $z = xy$ is sub-exponential with parameters $(\sqrt{2} \sigma_1 \sigma_2, \sqrt{2} \sigma_1 \sigma_2)$.}
\end{mylemma}
\begin{myproof}[(sketch)]
A direct computation of the moment-generating function of $z$ yields
\begin{equation}\label{E:proof_lemma_subexp_010}
    \mathbb{E}( e^{l z}) = \frac{1}{\sqrt{1-\sigma_1^2 \sigma_2^2 l^2}}.
\end{equation}
It is then not hard to verify that
\begin{equation}\label{E:proof_lemma_subexp_020}
    \frac{1}{\sqrt{1- \sigma_1^2 \sigma_2^2 l^2}} \leq \exp \left({\frac{2\sigma_1^2 \sigma_2^2 l^2}{2}}\right), 
\end{equation}
for all $l$ such that $|l|<1/(\sqrt{2}\sigma_1\sigma_2)$, and by combining \eqref{E:proof_lemma_subexp_010} and \eqref{E:proof_lemma_subexp_020} the statement of the lemma follows. In order to show inequality \eqref{E:proof_lemma_subexp_020}, one can begin by showing that $1/(1-w) \leq e^{2w}$ is valid for $0 \leq w \leq 1/2$, then apply the square root to the inequality and finally substitute $w = \sigma_1^2 \sigma_2^2 l^2$.
\end{myproof}

We will now show that $\bbS_M$ in \eqref{E:sample_covariance} and $\bbL$ are simultaneously diagonalizable, provided that $M$ is sufficiently large. For notational purposes, we define the matrix $\bbB^{(M)} := \bbV^\top \bbS_{M} \bbV$.
\begin{myproposition}\label{P:diagonal}
\emph{For $M\rightarrow \infty$, the eigenbasis $\bbV$ diagonalizes $\bbS_M$, i.e., for all $i \neq j$ it holds that
\begin{equation}\label{E:prop_diagon_010}
\lim_{M \to \infty} [\bbV^\top \bbS_{M} \bbV]_{ij} = \lim_{M \to \infty} B^{(M)}_{ij} = 0.
\end{equation}
}
\end{myproposition}
\vspace{2pt}
\begin{myproof}
\noindent By combining \eqref{E:redefinition_filter_H} and \eqref{E:sample_covariance} it follows that
\begin{equation}\label{E:proof_prop_diagonal_010}
\bbB^{(M)} = \frac{1}{M} \sum_{k=1}^M \bbLambda_k \bbV^\top \bbx_k \bbx_k^\top \bbV \bbLambda_k = \frac{1}{M} \sum_{k=1}^M \bbw_k \bbw_k^\top,
\end{equation}
where we have defined $\bbw_k := \bbLambda_k \bbV^\top \bbx_k$. Since $\bbx_k$ is a multivariate Gaussian random variable, it follows that ${\bbw_k~\sim\ccalN({\bf{0}}, \sigma^2 \bbLambda_{k}^2)}$. For each entry $i,j$ in \eqref{E:proof_prop_diagonal_010} we thus have
\begin{equation}\label{E:proof_prop_diagonal_020}
B^{(M)}_{ij} = \frac{1}{M} \sum_{k=1}^M [\bbw_k]_i [\bbw_k]_j,
\end{equation}
where $[\bbw_k]_i \sim \ccalN(0, \sigma^2 [\bbLambda^2_k]_{ii})$ and $[\bbw_k]_j \sim \ccalN(0, \sigma^2 [\bbLambda^2_k]_{jj})$ are independent. Lemma~\ref{L:subexponential} now implies that $[\bbw_k]_i [\bbw_k]_j$ is a sub-exponential random variable with parameters $(v_k, b_k):=(\sqrt{2} \sigma^2 [\bbLambda_k]_{ii} [\bbLambda_k]_{jj}, \sqrt{2} \sigma^2 [\bbLambda_k]_{ii} [\bbLambda_k]_{jj})$. Consequently, $B^{(M)}_{ij}$ is equal to the average of $M$ independent sub-exponential random variables, each of them having zero mean. Define then the coefficients
\begin{align}\label{E:proof_prop_diagonal_040}
v_*^2 := \frac{1}{M} \sum_{k=1}^M v^2_k,  \qquad b_* := \max_{k=1, \ldots, M} b_k.
\end{align}
The sub-exponential tail bound in \eqref{E:def_sub_exponential_bound} yields
\begin{equation}\label{E:proof_prop_diagonal_050}
\mathbb{P}( | B^{(M)}_{ij} | \geq l) \leq 2 \exp \left({\frac{-Ml^2}{2v_*^2}}\right),
\end{equation}
for $0 \leq l \leq v_*^2/b_*$. Recall that every $[\bbLambda_k]_{ii}$ is upper bounded by $1$, hence $v_*^2 \leq 2 \sigma^4$.
Consequently, for small enough $l>0$:
\begin{equation}\label{E:proof_prop_diagonal_060}
\lim_{M \to \infty} \mathbb{P}( | B^{(M)}_{ij} | \geq l) \leq \lim_{M \to \infty} 2 \exp \left({\frac{-Ml^2}{4 \sigma^4 }}\right) = 0,
\end{equation}
which proves the proposition.
\end{myproof}

Proposition~\ref{P:diagonal} guarantees that the eigenbasis $\bbV$ can be recovered by performing an eigendecomposition of $\bbS_M$ for large enough $M$. 
While in most practical instances of network inference $M$ will not be unbounded, Proposition~\ref{P:diagonal} nevertheless can be used as the basis for an inference algorithm even if only a finite number of observations are available; see Section~\ref{Ss:optimal_laplacian}. 

We remark that the validity of \eqref{E:prop_diagon_010} does not imply that $\lim_{M \to \infty}  \bbS_{M}$ exists. Indeed, our weak assumptions on $T_k$ and $\alpha_{t}^{(k)}$ -- which translate into mild conditions on $\bbH_k$ -- could result in an $\bbS_{M}$ that does not converge for increasing $M$. However, even if $\bbS_M$ does not converge to a \emph{specific} matrix, we may interpret \eqref{E:prop_diagon_010} at stating that $\bbS_M$ converges to the \emph{set} of matrices diagonalized by $\bbV$.

\subsection{Inferring the eigenvalues}\label{Ss:inferring_eigenvalues}
As will be shown in Proposition~\ref{P:eigenvalues}, $\bbS_M$ also provides insights about the eigenvalues $\bblambda$ of the unknown Laplacian $\bbL$. 
In proving this, we make use of the following lemma, whose proof is omitted for being analogous to that of Lemma~\ref{L:subexponential}.
\vspace{0.15cm}
\begin{mylemma}\label{L:subexponential_2}
\emph{
Let $x \sim \ccalN(0, \sigma^2)$, then the random variable $z = x^2$ is sub-exponential with parameters $(2 \sigma^2, 4 \sigma^2)$ and $\mathbb{E}(z) = \text{var}(x) = \sigma^2$.}
\end{mylemma}

\vspace{0.25cm}
\begin{myproposition}\label{P:eigenvalues}
\emph{
For every $\delta > 0$ there exists a large enough number of observations $M_\delta$ such that, for all $i < j$,
\begin{equation}\label{E:prop_eigenvalues_010}
B^{(M)}_{ii} > B^{(M)}_{jj},
\end{equation}
with probability at least $1-\delta$ for every $M \geq M_\delta$.}
\end{myproposition}
\begin{myproof}
\noindent Beginning from \eqref{E:proof_prop_diagonal_010}, it follows that
\begin{equation}\label{E:proof_eigenvalues_010}
B^{(M)}_{ii} = \frac{1}{M} \sum_{k=1}^M ([\bbw_k]_i)^2,
\end{equation}
where $[\bbw_k]_i \sim \ccalN(0, \sigma^2 [\bbLambda^2_k]_{ii})$. Invoking Lemma~\ref{L:subexponential_2}, we have that $([\bbw_k]_i)^2$ is a sub-exponential random variable with parameters $(v^i_k, b^i_k):=(2 \sigma^2 [\bbLambda^2_k]_{ii}, 4 \sigma^2 [\bbLambda^2_k]_{ii})$. Thus, $B^{(M)}_{ii}$ is equal to the average of $M$ independent sub-exponential random variables, each of which has mean $\sigma^2 [\bbLambda^2_k]_{ii}$ (Lemma~\ref{L:subexponential_2}). Define the global parameters $v_{*i}^2$ and $b_{*i}$ for each $i$ as
\begin{equation}\label{E:global_parameters_i}
v_{*i}^2 := \frac{1}{M} \sum_{k=1}^M (v^i_k)^2,  \qquad b_{*i} := \max_{k=1, \ldots, M} b^i_k,
\end{equation}
and the expected value 
\begin{equation}\label{E:def_mean_values}
    e_i := \mathbb{E}(B^{(M)}_{ii}) =  \frac{\sigma^2}{M} \sum_{k=1}^M  [\bbLambda^2_k]_{ii}
\end{equation}
Then, we can again leverage the sub-exponential tail bounds in \eqref{E:def_sub_exponential_bound} to obtain
\begin{equation}\label{E:proof_eigenvalues_020}
\mathbb{P}\!\left( \left| B^{(M)}_{ii} - {e}_i \right| \!\geq\! l \!\right) \!\leq \!2 \exp\!\left(\!{\frac{-Ml^2}{2v_{*i}^2}}\!\right) \!\leq \!2 \exp \!\left(\!{\frac{-Ml^2}{8 \sigma^4}}\!\right)\!, 
\end{equation}
for $0 \leq l \leq v_{*i}^2/b_{*i}$. The last inequality follows from the fact that $\bbLambda_k$ is upper bounded by $1$ for all $k$, which results in the bound $v_{*i}^2 \leq 4 \sigma^4$. A direct application of the union bound on \eqref{E:proof_eigenvalues_020} yields
\begin{equation}\label{E:proof_eigenvalues_030}
\mathbb{P}\left( \bigcup_{i=1}^N  \left| B^{(M)}_{ii} - {e}_i \right| \geq l \right) \leq 2 N \exp \left({\frac{-Ml^2}{8 \sigma^4}}\right), 
\end{equation}
for $0\leq l \leq v_{*}^2/b_{*} := \min_i (v_{*i}^2/b_{*i})$, from which it immediately follows that
\begin{equation}\label{E:proof_eigenvalues_040}
\mathbb{P}\left( \bigcap_{i=1}^N  \left| B^{(M)}_{ii} - {e}_i \right| < l \right) \geq 1-  2 N \exp \left({\frac{-M l^2}{8 \sigma^4}}\right) \geq 1-\delta, 
\end{equation}
where we fixed a desired probability level at $1-\delta$. 
Our goal now is to choose $l$ small enough to ensure that \eqref{E:prop_eigenvalues_010} is satisfied and then solve for the corresponding number of observations $M_\delta$ in \eqref{E:proof_eigenvalues_040} using such an $l$. To do this, first recall that the eigenvalues of the Laplacian in $\bbLambda$ satisfy the ordering $0 = \Lambda_{11} < \ldots <\Lambda_{NN}$. 
Hence, we know that the diagonal entries of $\bbLambda_k$ will be inversely sorted [cf.~\eqref{E:redefinition_filter_H}], i.e., $1= [\bbLambda_k]_{11} > \ldots > [\bbLambda_k]_{NN}$. 
We further assume that $[\bbLambda_k]_{ii} > [\bbLambda_k]_{jj} + \tau$ when $i<j$ for some $\tau > 0$, where $\tau$ does not depend on $M$.
It then follows from \eqref{E:def_mean_values} that ${e}_i > {e}_j + \sigma^2 \tau^2$ for $i<j$. Consider a deviation from the mean $l^*:=  \sigma^2 \tau^2/\beta$ where $\beta \geq 2$ is large enough to ensure that $l^* \leq v_{*}^2/b_{*}$. By specializing \eqref{E:proof_eigenvalues_040} to $l=l^*$ and solving for $M$ as a function of $\delta$, we have that for all $M$ such that
\begin{equation}\label{E:proof_eigenvalues_050}
M \geq M_\delta := \frac{8 \beta^2}{\tau^4} \log\left(\frac{2N}{\delta}\right),
\end{equation}
every random variable $B^{(M)}_{ii}$ is at most at a distance $l^*$ from its mean with probability at least $1-\delta$. Since by definition $l^* < ({e}_i - {e}_j)/2$ for $i<j$, this means that the variables $B^{(M)}_{ii}$ are sorted in the same order as their means with high probability, i.e., $B^{(M)}_{ii} > B^{(M)}_{jj}$ for $i <j$ with probability at least $1-\delta$, as we wanted to show.
\end{myproof}

In Section~\ref{Ss:inferring_eigenvectors} we discussed that $\bbS_M$ need not converge for large $M$. Nevertheless, \eqref{E:prop_eigenvalues_010} is stating that, even in the diverging case, the diagonal elements of $\bbB^{(M)} =\bbV^\top \bbS_{M} \bbV$ follow a specific order with high probability. 
This observation, in combination with Proposition~\ref{P:diagonal}, is leveraged in Section~\ref{Ss:optimal_laplacian} to develop a network topology inference algorithm for finite $M$.

\subsection{Recovering the optimal Laplacian matrix}\label{Ss:optimal_laplacian}

As discussed earlier, selecting a Laplacian $\bbL$ that is consistent with the observations $\{\bby_k\}$ is in general an under-determined problem. Even when fixing the eigenbasis $\bbV$ and the ordering of the eigenvalues, there is freedom in choosing the exact eigenvalues as long as the order is preserved. 
Consequently, we seek to recover an \emph{optimal} $\bbL$ among all those consistent with the observed data. Our notion of optimality is based on sparsity, but other features might be selected as well.

Denoting by $\bbS_M = \tilde{\bbV} \tilde{\bbLambda} \tilde{\bbV}^\top$ the eigendecomposition of $\bbS_M$ where the eigenvalues are sorted in increasing order, our inferred Laplacian ${\bbL}^*$ can be found as the solution of the following convex optimization problem.
\begin{subequations}\label{E:problem_eqall}
\begin{align}
\{ {\bbL}^*, \tilde{\bbL}^*, &{\bblambda}^*\} := \argmin_{\{ \bbJ,  \bbK, \bbbeta\}} \| \bbJ \|_1 \label{E:problem_eq1}\\
\text{subject to}& \quad  J_{ij} = J_{ji} \leq 0 \,\, \text{for} \,\, i \neq j, \quad \bbJ \mathbf{1} = \mathbf{0}, \label{E:problem_eq2}\\
& \quad {\bbK} = \tilde{\bbV} \diag(\bbbeta) \tilde{\bbV}^\top, \quad \| \bbJ - \bbK \|_\mathrm{F} \leq \epsilon_1, \label{E:problem_eq3}\\
& \quad  \beta_i \geq \beta_{i+\eta} + \epsilon_2 \,\, \text{for} \,\, i = 1, \ldots, N-\eta. \label{E:problem_eq4}
\end{align}
\end{subequations}
Since the elementwise $\ell_1$ norm $\| \bbJ \|_1 := \sum_{ij} |J_{ij}|$ is simply a convex relaxation of the $\ell_0$ (pseudo-)norm, the objective \eqref{E:problem_eq1} promotes sparsity in $\bbL^*$, i.e., the optimal choice for $\bbJ$. Alternatively, the $\ell_1$ norm can be replaced by its iterative reweighted counterpart~\cite{Candes2008}, which has shown to perform better in practice. 
The constraints in \eqref{E:problem_eq2} force the output to be a valid Laplacian, namely, $\bbL^*$ must have non-positive off-diagonal elements and each row must sum up to zero. 
Notice that these two requirements enforce diagonal dominance of $\bbL^*$ which, in turn, ensures positive semi-definiteness. 
The constraints in \eqref{E:problem_eq3} impose that $\bbL^*$ must be close to being diagonalized by $\tilde{\bbV}$. 
It was shown in Section~\ref{Ss:inferring_eigenvectors} that $\tilde{\bbV}$ coincides with $\bbV$ for arbitrarily large number of observations $M$. 
However, for all practical implementations, $M$ is finite and thus, we do not require $\bbL^*$ to be diagonalized by $\tilde{\bbV}$ directly.
Rather, we require our output $\bbL^*$ to be close (as measured by the Frobenius norm) to another matrix $\tilde{\bbL}^*$ (the optimal $\bbK$) that is diagonalized by $\tilde{\bbV}$. 
Note that in practise the matrix variable $\bbK$ does not need to be constructed, but $\tilde{\bbV} \diag(\bbbeta) \tilde{\bbV}^\top$ can be directedly substituted into the norm.
We remark further that, we could replace the Froebenius norm here by the maximum, thereby reducing the problem to a linear program.
Lastly, \eqref{E:problem_eq4} incorporates the fact that the eigenvalues of $\bbS_M$ and the true Laplacian are inversely ordered by forcing this inverse order to $\tilde{\bbL}^*$. 
Here $\eta$ is a positive integer that we can choose.
Notice that when $\eta = 1$ we impose a strict order on the eigenvalues whereas for $\eta > 1$ we impose a laxer order for the cases in which $M$ is not large enough. 
Finally, the constant $\epsilon_2 >0$ can be chosen freely, since it will only vary the scale of the recovered Laplacian $\bbL^*$. 
Notice that this scale ambiguity is unsurmountable given that in \eqref{E:final_multiple_consensus} a common factor across all (unknown) $\alpha_{t}^{(k)}$ can be absorbed into the unknown $\bbL$.

In a nutshell, given a series of observations $\{ \bby_k \}_{k=1}^M$ following model \eqref{E:final_multiple_consensus}, we propose to recover (a scaled version of) $\bbL$ by first constructing $\bbS_M$ as in \eqref{E:sample_covariance} to extract its eigenbasis $\tilde{\bbV}$, and then solving problem \eqref{E:problem_eqall}. 
In the next section we assess the practical performance of this approach.

\begin{figure*}
    \centering	
    \begin{subfigure}{.25\textwidth}
        \centering
        \includegraphics[width=\textwidth, height=0.9\textwidth]{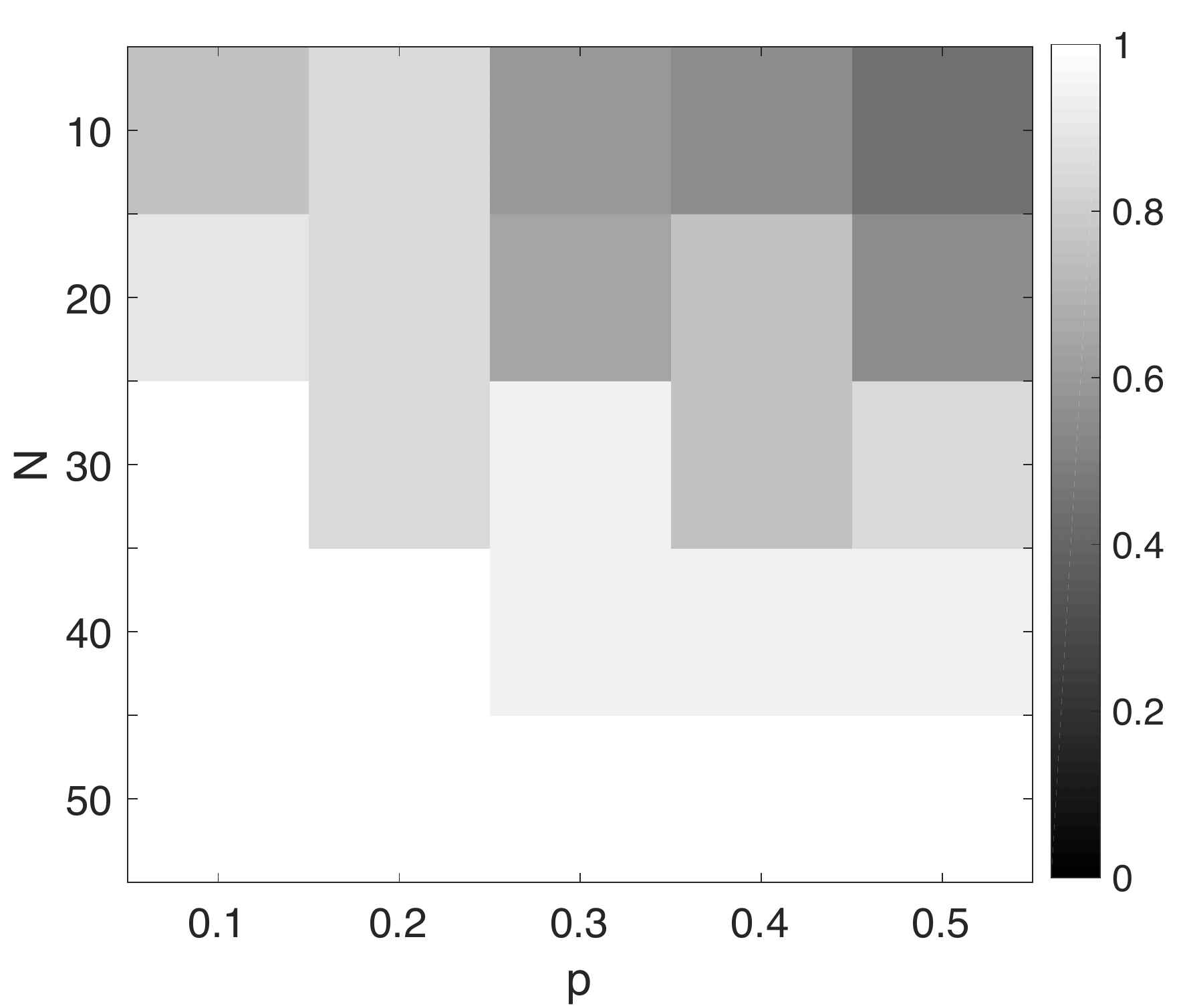}
        \caption{}
                \label{fig:sub1}
    \end{subfigure}%
    \hspace{0.2cm}
    \begin{subfigure}{.33\textwidth}
        \centering
        \includegraphics[width=\textwidth, height=0.68\textwidth]{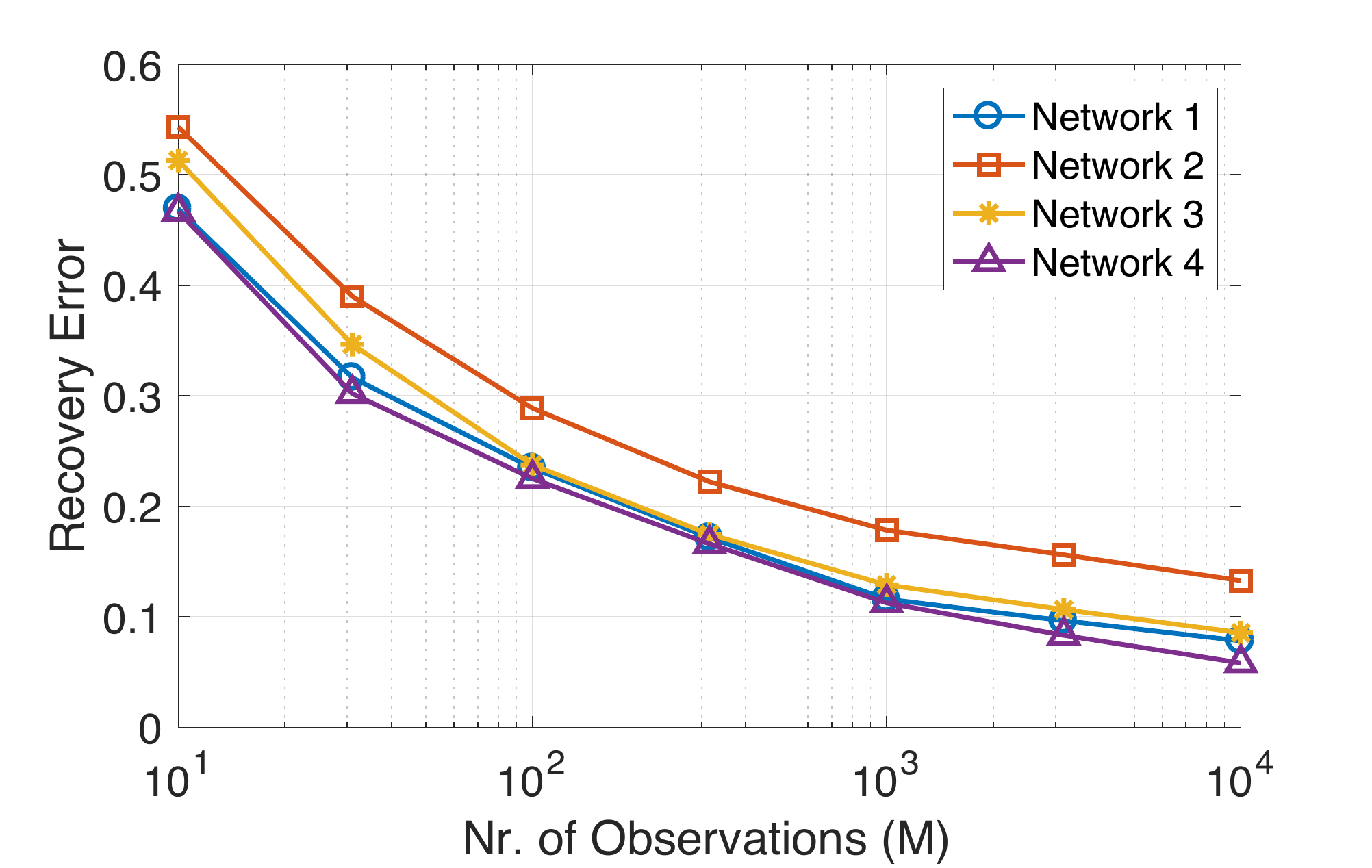}
        \caption{}
        \label{fig:sub2}
    \end{subfigure}
        \hspace{0.2cm}
    \begin{subfigure}{.37\textwidth}
        \centering
        \includegraphics[width=\textwidth, height=0.61\textwidth]{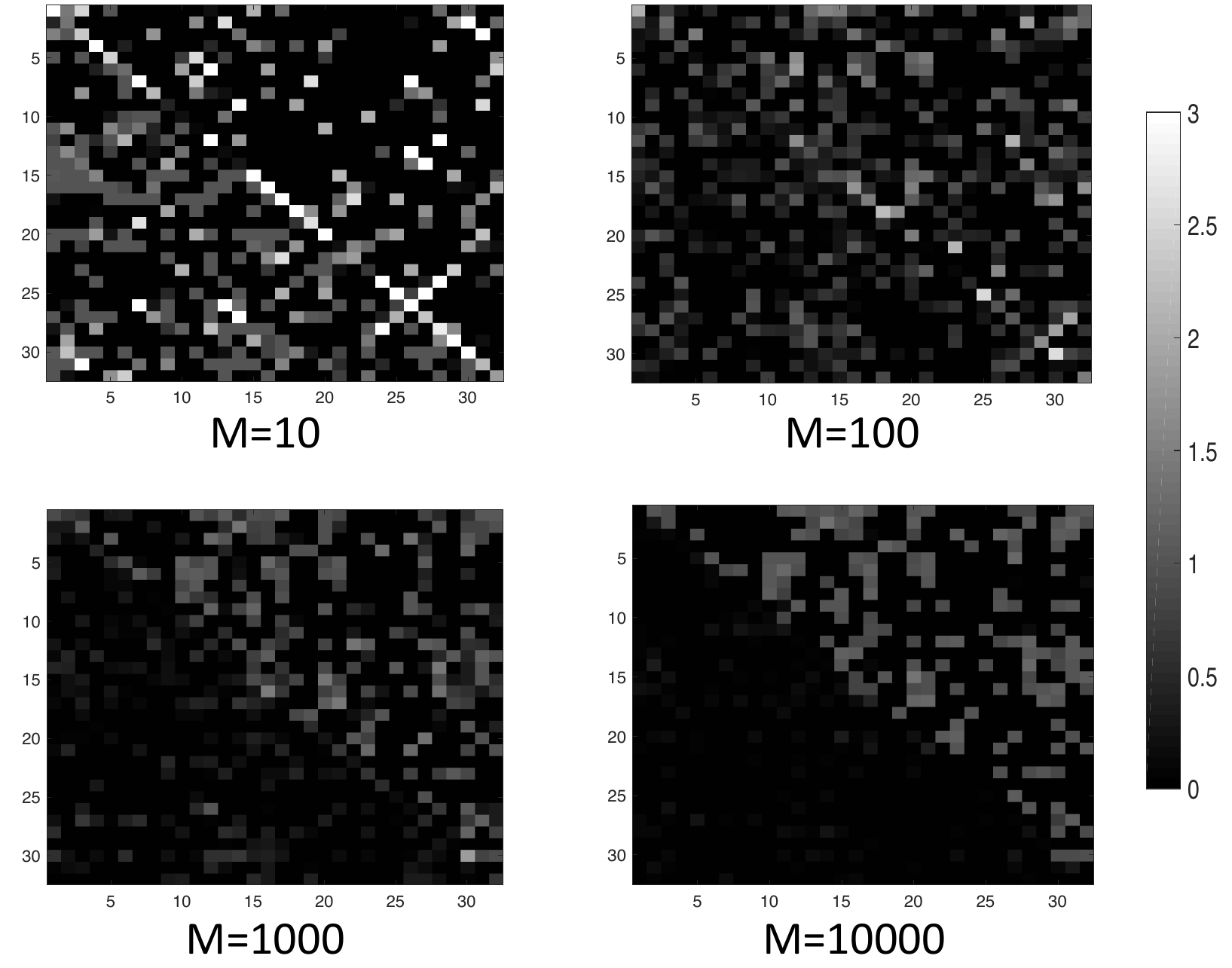}
        \caption{}
        \label{fig:sub3}
    \end{subfigure}%
    \caption{(a) Rate of recovery for Erd\H{o}s-R\'enyi graphs as a function of the number of nodes $N$ and the probability of edge appearance $p$. (b) Error in recovering four social networks as a function of the number of opinion profiles $M$ observed. (c) Examples of the recovered Laplacians for $M \in \{10, 10^2, 10^3, 10^4\}$. Upper triangular parts of matrices show the recovered edges and lower triangular parts (and diagonals) show discrepancy with true Laplacian.}
    \label{F:numerical_experiments}
\end{figure*}

\section{NUMERICAL EXPERIMENTS}\label{S:numerical_experiments}

We present two test cases, where our goal is to recover synthetic and real-world networks, respectively.

\vspace{0.1cm}
\noindent {\bf Erd\H{os}-R\'enyi networks.} To validate our method, we test its performance when the eigenbasis $\bbV$ is perfectly known. Intuitively, this situation will arise when an infinite number of observations $M$ is available (cf. Proposition~\ref{P:diagonal}). 
Clearly, a high recovery rate under this setting is a necessary condition for acceptable recovery in the finite $M$ regime. Hence, we consider Erd\H{o}s-R\'enyi (ER) random graphs \cite{bollobas1998random} of varying sizes $N \in \{10, \ldots , 50\}$ and edge probabilities $p \in \{0.1, \dots , 0.5\}$. For each combination of $N$ and $p$ we generate $20$ networks, compute their associated Laplacian $\bbL = \bbV \bbLambda \bbV^\top$, and then try to recover it by solving \eqref{E:problem_eqall} for $\tilde{\bbV} = \bbV$, $\eta = 1$, and $\epsilon_1 = 0$. Recall that the choice of $\epsilon_2$ only affects the scale of the recovered Laplacian and, thus, is inconsequential to the recovery performance. For every network generated, we consider the recovery to be successful if the error $\| \bbL^* - \bbL \|_{\mathrm{F}} / \| \bbL \|_{\mathrm{F}}$ is less than $2 \times 10^{-2}$. Fig.~\ref{fig:sub1} portrays the recovery rates (averaged across the $20$ realizations) as a function of $N$ and $p$. 

We first observe that the recovery rates are high. The overall recovery rate is $0.85$ and, if we increase the threshold for success recovery to $5 \times 10^{-2}$, this rate becomes $0.97$. Secondly, we note that as $N$ increases, recovery becomes almost certain. 
The reason for this is that, after assuming perfect knowledge of $\bbV$, having two sparse Laplacians that share identical sets of eigenvectors becomes less probable for larger $N$. Finally, we see a decay in performance for increasing $p$, which can be attributed to the fact that we are imposing sparsity on the recovered Laplacian even for relatively large values of $p$.

\vspace{0.1cm}
\noindent {\bf Real-world social networks.} 
Consider four social networks defined on a common set of $N=32$ nodes, which represent students from the University of Ljubljana\footnote{Access to the data and additional details are available at \urlstyle{same}\url{http://vladowiki.fmf.uni-lj.si/doku.php?id=pajek:data:pajek:students}}.
Edges in each of the networks represent different types of interactions among the students, and were constructed by asking each student to select a group of preferred college mates for different situations, e.g., to discuss a personal issue or to invite to a birthday party. The considered networks are unweighted and symmetric, and the edge between nodes $i$ and $j$ exists if either student $i$ picked $j$ in the questionnaire or vice versa.

For each of the four networks, our goal is to recover the true Laplacian $\bbL$ from the observation of $M$ synthetic consensus dynamics by solving \eqref{E:problem_eqall}, where we vary $M$ from $10$ to $10^4$; see Fig.~\ref{fig:sub2}. 
We consider the same metric for recovery error as in the previous experiment, averaged over $20$ realizations of a synthetic dynamics.
This synthetic dynamics was generated by drawing the input $\bbx$ from a standard multivariate Gaussian distribution, selecting $T_k$ uniformly at random in $\{3, 4, 5\}$ and each rate $\alpha_t^{(k)}$ uniformly at random in $(0, 1/\lambda_{\mathrm{max}}(\bbL))$. In solving \eqref{E:problem_eqall} we set $\epsilon_1$ equal to the smallest possible value (found via iterative search) that guarantees feasibility of \eqref{E:problem_eqall} and $\eta = 5$, although recovery was robust to the specific value chosen for $\eta$.

As displayed in Fig.~\ref{fig:sub2}, we observe a monotonous decrease of the recovery error with increasing $M$ for all networks. This is not surprising since we know that for larger $M$, the sample eigenbasis $\tilde{\bbV}$ becomes closer to the real eigenbasis $\bbV$ and thereby facilitates recovery. 
In Fig.~\ref{fig:sub3} we show specific instances of $\bbL^*$ corresponding to Network 1 for different number of observations $M$. 
To facilitate interpretation, the matrices shown in the figure are split along the diagonal: the upper triangular portion of each matrix (excluding the diagonal) corresponds to the absolute values of the entries of the recovered $\bbL^*$, whereas the lower triangular portion and the diagonal correspond to the difference with the true Laplacian, i.e. entries of $|\bbL- \bbL^*|$. 
As expected, the discrepancy with the real Laplacian (lower triangle) decreases with increasing $M$. 
Moreover, note that we here consider a weighted network recovery.
If we are only interested in recovering the support of the graph, then we can do better even with only $M=100$ samples by post-processing our results. More precisely, if we only keep the $155$ (true number of edges in $\bbL$) strongest links in $\bbL^*$, $118$ of them coincide with the edges present in $\bbL$. This overlap increases to $149$ and $155$ for $M=10^3$ and $M=10^4$, respectively. Notice that even for this last case the error in Fig.~\ref{fig:sub2} is not $0$, due to small differences in the actual weights of the recovered edges.

\section{CONCLUSION}\label{S:conclusion}
We have proposed a novel technique for the identification of a network based on observing snapshots of a number of independent consensus processes of unknown duration.
To achieve this, we formulated a convex optimization problem that outputs a sparse, valid Laplacian which is provably consistent with the spectral information obtained from the consensus observations. 
 
Our results pave the way for several interesting avenues of future work including:
(i) investigation of the trade-off between specific network topologies and the required sample size $M$ to achieve a desired level of recovery performance; 
(ii) consideration of a richer class of dynamical models, including non-deterministic processes such as switched systems; and 
(iii) extensions of the proposed algorithm that incorporate generative random network models as priors in the network inference problem.


\bibliographystyle{IEEEtran}
%
\bibliography{citations}

\end{document}